\documentclass[twocolumn,superscriptaddress,showpacs,preprintnumbers,amsmath,amssymb,pre]{revtex4}
\usepackage{graphicx}
\usepackage{dcolumn,xcolor,ulem}
\usepackage{amsmath} 
\usepackage{amssymb}
\usepackage{amsfonts}
\usepackage{bm}
\usepackage[latin1]{inputenc}
\usepackage{hyperref} 

\begin{document}

\title{Overcooked agar solutions:\\
impact on the structural and mechanical properties of agar gels}

\author{Bosi Mao}
\author{Ahmed Bentaleb}
\author{Fr\'ed\'eric Louerat}
\author{Thibaut Divoux}
\email[]{divoux@crpp-bordeaux.cnrs.fr}
\author{Patrick Snabre}
\email[]{snabre@crpp-bordeaux.cnrs.fr}
\affiliation{Université Bordeaux, Centre de Recherche Paul Pascal, UPR~8641, 115 av. Dr. Schweitzer, 33600 Pessac, France}

\date{\today}

\begin{abstract}
Thermoreversible hydrogels are commonly prepared by cooling down to ambient temperature, aqueous polymer solutions first brought to a boil. The incubation time of the polymer solution at such a high temperature is traditionally kept to a minimum to minimize its impact on the subsequent gelation. Here we study the effect of a prolonged heating of a 1.5\% w/w agar solution at $T=80^{\circ}$C, well above the gelling temperature. The incubation time $\mathcal{T}$ of the polysaccharide solution is varied from a few hours up to five days. We show that the agar solution ages as the result of both the hydrolysis and the intramolecular oxidation of the polysaccharides. As a consequence, both the viscosity and the pH of the solution decrease continuously during the incubation period. Furthermore, samples withdrew at different incubation times are cooled down to form gels which structure and mechanical properties are systematically determined. Cryoelectron microscopy and X-ray diffraction experiments reveal that agar gels formed from solutions of increasing incubation times, display a coarser microstructure composed of micron-sized foils which result from the condensation of the polysaccharides and contrast with the fibrous-like microstructure of gels prepared from a fresh agar solution. Along with structural changes, a prolonged incubation time of the polymer solution at $T=80^{\circ}$C leads to weaker agar gels of significantly lower shear elastic modulus. Moreover, extensive macro-indentation experiments coupled to direct visualization show that increasing the incubation time of the agar solution up to a few days decreases the yield strain of the gel by a factor of three, while the rupture scenario turns continuously from brittle to ductile-like. Our study suggests that the incubation time of agar solutions at high temperature could be used as an external control parameter to finely tune the mechanical properties of agar-based gels.
\end{abstract}


\maketitle

\section{Introduction}

Biopolymer gels encompass a wide array of soft materials currently used in countless application fields ranging from biomedical domains, e.g. cell encapsulation \cite{Hunt:2010}, tissue engineering \cite{Lee:2001,VanVlierberghe:2011}, etc. to food science \cite{Mezzenga:2005} and molecular gastronomy \cite{Barham:2010}. Such gels are made of proteins such as casein, albumine, $\beta$-lactoglobilin, etc. and/or polysaccharides such as alginate, carrageenan and agarose, etc. The gelation scenario of these materials is induced by the addition of co-solutes such as salts, by changes in pH or ionic strength, or by physical or thermal means leading to a conformational change of the polymer chains during their self-assembly \cite{Clark:1987}. By tuning the gel components and/or the gelation path, one can build soft solids with a wide array of mechanical properties in view of specific applications \cite{Rinaudo:2008,Calvert:2009}. 
The search for improving hydrogels mechanical strength, and develop tough and highly stretchable materials has produced a wealth of crosslinking methods mainly based on the use of a double networks, or fancy crosslinkers \cite{Gong:2003,Sun:2012,Zhao:2014,Lin:2014,Grindy:2015}. However, such methods may not always be adapted for biomedical applications for which the use of natural polymers alone in water is often preferred to modify the gel properties.    
Here we discuss an alternate route for engineering specifically the mechanical properties of polysaccharide-based hydrogels without external crosslinkers. The method is based on the intermolecular condensation of the polysaccharide chains, which can be controlled by the incubation time of the polymer solution above the gelation temperature. Intermolecular condensation makes it possible to coarsen the hydrogel microstructure and soften the mechanical properties as extensively illustrated on agar gels in the present article.  

\begin{figure*}[!ht]
\centering
	\includegraphics[width=0.9\linewidth]{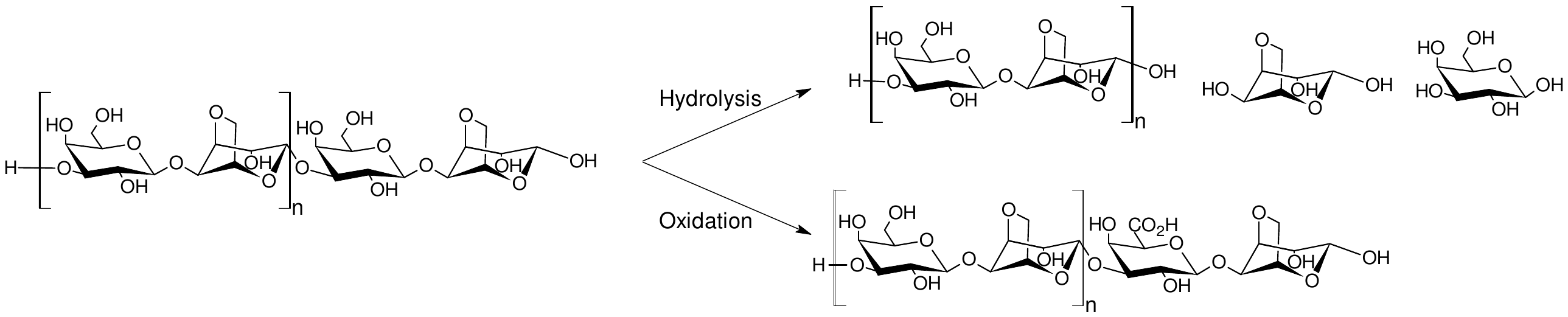}
\caption{(Left) Illustration of the disaccharide unit (AB)$_n$ composing the agarose molecule. (Right) Product of the hydrolysis and oxidation of the agarose molecule as discussed in section~\ref{Solaging}.       
\label{fig0}}
\end{figure*} 

Agar is a mixture of marine algal polysaccharides among which the gelling component agarose is used as a thickener in food products, and once gelled, plays a key role in numerous biotechnological applications ranging from growth medium for microorganisms to substitute for biological tissues \cite{Beruto:1999,Rinaudo:2008,Culjat:2010}. Agarose is only soluble in boiling water and the gelation occurs via hydrogen bonds \cite{Tako:1988,Braudo:1992} upon cooling the sol below about 40$^{\circ}$C, leading to a thermoreversible gel which, in turn, only melts above 80$^{\circ}$C \cite{Nijenhuis:1997b}.
If agar gels have attracted a lot of attention over the passed 30 years, especially the sol-gel transition scenario which involves a subtle interplay between spinodal demixing and direct gelation \cite{Feke:1974,SanBiagio:1996,Manno:1999,Matsuo:2002}, paradoxically enough, the role of the incubation time of the agar sol at high temperature prior to the gel formation has been poorly investigated. Furthermore, the quantitative influence of an extended heating of the agar sol on the mechanical properties of the subsequent gel remains unknown. Although increasing the incubation time of the agar sol up to a few hours has shown no impact on the gel mechanical properties \cite{Whyte:1984}, agar sols are maintained at large temperatures for hours if not days in numerous industrial applications which motivates a thorough study of the impact of the incubation time on agar gels over long durations.  

The goal of the present manuscript is twofold. First we report the evolution of the properties of an agar sol that is maintained at $T=80^{\circ}$C from a few hours up to 5 days. The decrease of both the pH and the sol intrinsic viscosity leads us to conclude to the hydrolysis and the oxidation of the polysaccharides in solution. Second, samples drawn at regular time intervals are used to prepare gels which structure and mechanical properties are systematically investigated. Cryo-SEM and X-ray diffraction spectroscopy reveal that gels formed after prolonged incubation time of the agar sol display a coarser microstructure composed of micron-sized foils. Small amplitude oscillatory shear experiments and time resolved monitoring of crack growth during macro-indentation tests further show that the latter gels display weaker elastic properties, a smaller yield strain and a more ductile failure scenario than gels formed from a fresh agar sol. The present work offers an extensive description of the evolution of the gel mechanical properties associated with the prolonged heating of agar sol and shows that incubation of agar sol at large temperature could be used to tune the mechanical properties of agar gels in a reproducible fashion.

\begin{figure*}[!t]
\centering
	\includegraphics[width=0.9\linewidth]{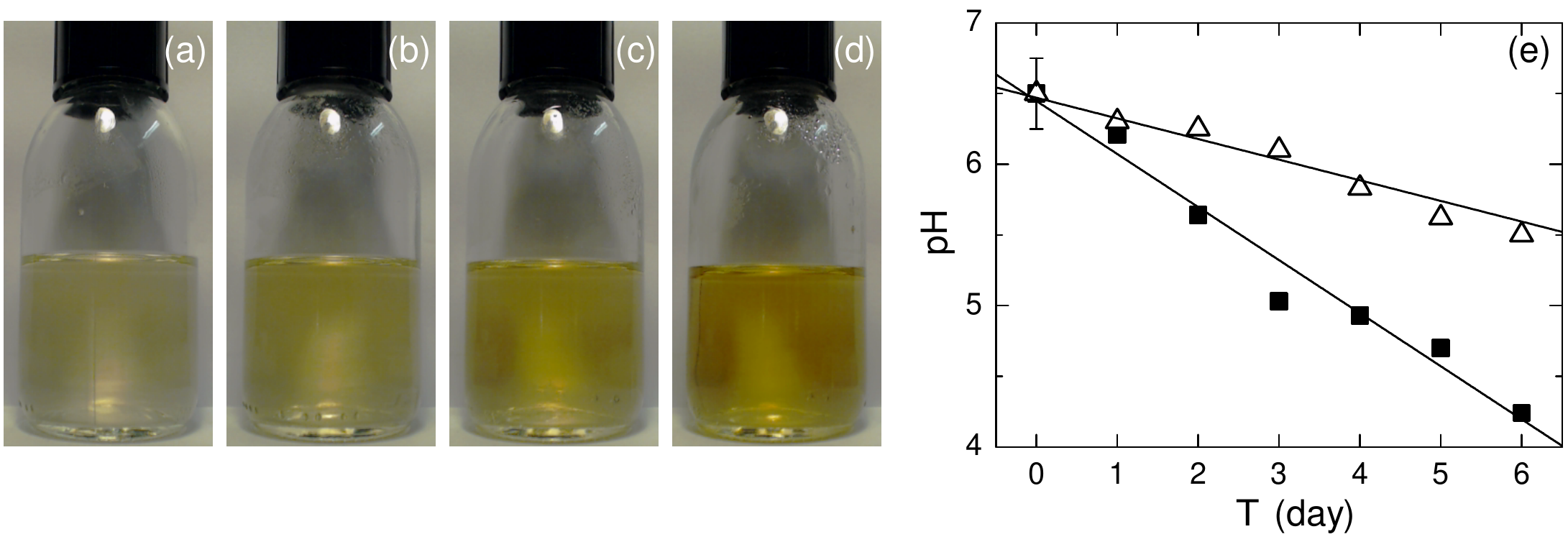}
\caption{(a)-(d) Pictures of the agar sol after different incubation times $\mathcal{T}$, respectively from left to right: $\mathcal{T}=$ a few hours, 3 days, 6 days and 12 days. The scale is fixed by the 4.5~cm diameter of the glass bottle. (e) Evolution of the pH measured at $T=80^{\circ}$C vs. the incubation time  $\mathcal{T}$ of the agar sol when the latter is either in contact with ambient air ($\blacksquare$) or in contact with a nitrogen atmosphere ($\bigtriangleup $). The black lines correspond in both cases to the best linear fit of the data: ${\rm pH}=6.5-0.15\mathcal{T}$ and ${\rm pH}=6.4-0.37\mathcal{T}$.      
\label{fig.sol}}
\end{figure*} 

\section{Experimental section}
 \label{ES}

 \subsection{Preparation of agar solutions} \label{Sample}
Samples of 200~mL are prepared by mixing 1.5\% w/w of agar powder for molecular biology composed at 70\% of agarose and 30\% of agaropectin (CAS 9002-18-0, ref.~A5306 Sigma-Aldrich) with milli-Q water (17~M$\Omega$.cm at 25$^{\circ}$C) brought to a boil. The temperature is kept at 100$^{\circ}$C for 10~min before being decreased to 80$^{\circ}$C. The agar sol is then introduced in a programmable testing chamber (Binder MK53) and maintained a constant temperature of ($80 \pm 1$)$^{\circ}$C. That moment sets the origin of time $\mathcal{T}=0$ and all the experiments performed later on with the solution are timestamped with respect to that date. Samples are extracted every day up to $\mathcal{T}=5$~days to monitor the evolution of the sol properties, which are summarized in section~\ref{Solaging}. Moreover, gels are prepared from the agar sol at regular time intervals following different protocols detailed in the remainder of section~\ref{ES}. Gels mechanical properties are determined as a function of the incubation time $\mathcal{T}$, as detailed below in section~\ref{Gelcharac}. We shall emphasize that the experiments reported in the manuscript have been repeated at least three times and tested on other brands of agar than Sigma-Aldrich, as discussed in the appendix.  

Let us recall here that agarose is a polymer composed of alternating disaccharide units (AB)$_n$, where A and B respectively denotes $\beta$-1,3 linked D-galactose and $\alpha$-1,4 linked 3,6-anhydro-L-galactose residues \cite{Araki:1956,Clark:1987,Matsuhashi:1990} [Fig.~\ref{fig0} (Left)]. Gels made of agar exhibit a porous microstruture filled with water and composed of bundles of about 0.1~nm diameter \cite{Whytock:1991} consisting of non-covalently crosslinked polysaccharides. The pores show a broad size distribution, up to a few hundred nanometers \cite{Brigham:1994,Chui:1995,Pernodet:1997,Xiong:2005,Rahbani:2013}. As a consequence, agar gels behave as viscoelastic soft solids which mechanical properties are mainly governed by the amount of agarose introduced: both the shear and compression elastic modulus $G'$ of agar gels scale as power laws of the agarose concentration \cite{Ramzi:1998,Normand:2000}. The thermal history upon cooling only impacts the mechanical properties of the gel for agarose concentrations larger than 2\% w/w \cite{Aymard:2001,Mao:2015a}, while the detailed nature of the polymer backbone, and especially the ester sulphate content, affects the water holding capacity of the gel \cite{Matsuhashi:1990}. Moreover, being mainly composed of water, agar gels are sensitive to water evaporation which leads, over long duration, to the gel shrinkage and results in a delayed detachment dynamics when the gel is casted in a Petri dish \cite{Divoux:2015}. Finally, agar gels display a brittle-like failure scenario \cite{Oates:1993} that can be delayed when explored under external stress \cite{Bonn:1998b,Bostwick:2013}: fractures grow perpendicularly to shear beyond a critical yield strain which value is governed by the molecular weight of the polymer \cite{Normand:2000}.   

\subsection{Experimental techniques} 
 \label{ET}

 \subsubsection{Rheology and macro-indentation} \label{Rheol}
 The viscosity $\eta$ of the agar sol is measured at regular time intervals $\mathcal{T}$ in a cone-and-plate geometry driven by a stress-controlled rheometer (DHR-2, TA instruments). The cone (diameter 60~mm, angle 2$^{\circ}$) is made of stainless steel and the bottom plate consists in a Teflon coated Peltier unit which allows us to control the temperature of the sample. Viscosity measurements are performed at $T=50^{\circ}$C and the geometry is pre-heated before being filled by the agar liquid solution. 

Gelation experiments are also performed at regular time intervals $\mathcal{T}$ in a parallel-plate geometry driven by a stress controlled rheometer (DHR-2, TA instruments). The upper plate (diameter $2r=40$~mm) is made of passivated duralumin and displays a surface roughness of ($4 \pm 2$)~$\mu$m as determined by Atomic Force Microscopy. The bottom plate consists in a smooth Teflon coated Peltier unit which allows us to impose a decreasing ramp of temperature from $T=70^{\circ}$C down to $20^{\circ}$C, at $\dot{\rm T}=1^{\circ}$C/min, leading to the gelation of the agar solution below about 36$^{\circ}$C. Evaporation is minimized by using a solvent trap filled with deionized water. Gelation experiments are performed under controlled normal force equals to zero (and not at constant gap width) to prevent any strain hardening of the gel, as explained in detail in the following reference \cite{Mao:2015a}. The decrease of the gap width compensates for the small and yet non-negligible contraction of the sample during the gelation.
The elastic and viscous moduli $G'$ and $G''$, are measured through small amplitude oscillations with a frequency $f=1$~Hz. To optimize the linear measurements, the strain amplitude $\gamma$ is adapted to the physical state of the sample: we apply $\gamma=1$~\% for $G'<1$~Pa, $\gamma=0.1$\% for $1<G'\leq 10$~Pa and $\gamma=0.01$\% for $G' >10$~Pa. Imposing a large strain amplitude while the sample is still liquid provides a better resolution to determine the viscoelastic moduli \cite{Ewoldt:2015}, and decreasing the strain amplitude as the sample becomes a gel allows us to remain within the linear regime and further prevents any debonding between the gel and the plates \cite{Mao:2015a}. 
 
Finally, macro-indentation experiments are conducted on agar gels prepared after different incubation times $\mathcal{T}$ of the agar sol. Agar gels shaped as flat cylinders of ($3.9 \pm 0.2$)~mm thick are prepared in glass Petri dishes of 50~mm diameter and left to gelify at room temperature, i.e. $T=(22\pm 2)^{\circ}$C. The latter temperature does not play a key role since the cooling rate has no influence upon neither the structure nor on the elastic properties of a 1.5\% agar gel \cite{Mao:2015a}. Once the gel is formed, a disk of $2R=34$~mm diameter is prepared by stamping the center of the dish with a circular punch tool, and removing the gel surrounding the disk. The untouched disk together with the original glass plate are placed on a transparent base in PMMA for the observation of the gel disk from below, during the subsequent indentation. The latter test is performed by means of a duralumin cylinder connected at one end to a stress-controlled rheometer (DHR-2, TA instruments) and ending with a circular flat surface on the other side (diameter 10~mm). The normal force sensor of the rheometer is used as a force gauge to determine the stress-strain $\sigma_N(\epsilon)$ relation during the sample indentation. In practice, the indenter is lowered towards the center of the gel disk at a continuous velocity v=100~$\mu$m/s, down to a position corresponding to a strain $\epsilon=50$\%. A camera (Logitech HD C920) and a flat mirror placed at an angle of $45^{\circ}$ with respect to the transparent PMMA plate are used to monitor the gel from below, which allows us to record the formation and the propagation of cracks that may occur at large enough strains as detailed in section~\ref{macroindent}.
 
 \subsubsection{Cryo scanning electron microscopy} \label{SEM}
 The microstructure of agar gels prepared from samples extracted of the same solution after different incubation times $\mathcal{T}$ at 80$^{\circ}$C is investigated by cryo-Scanning Electron Microscopy (cryo-SEM). Agar gels are prepared in Petri dishes left to cool at room temperature, i.e. $T=(22\pm 2)^{\circ}$C. A sample is taken with a scalpel and fixed to a metallic pin stub. Both are immersed in a liquid nitrogen bath for about 5~min. The sample is then placed in the preparation chamber of a SEM microscope (JEOL 6700F) which is cooled down to $T=-90^{\circ}$C. The sample is beheaded in situ and the temperature is increased at about $5^{\circ}$C/min up to $T=-50^{\circ}$C to allow the sublimation of the water enclosed inside the sample. After 5~min, the temperature is decreased back to $T=-85^{\circ}$C and the sample is coated with a nanolayer of gold-palladium. Finally, the sample is cooled down to $T=-160^{\circ}$C and introduced inside the observation chamber of the microscope. Images are taken in SEI mode at 5~kV.

 \subsubsection{X-ray experiments}\label{Xray}
X-ray diffraction spectra are obtained from dried gels prepared as follows: agar gels are casted in Crystal Polystyrene Petri dishes (diameter 55~mm), using samples extracted at different ages $\mathcal{T}$ from the solution maintained at 80$^{\circ}$C.  Gels are then left to dry for 3 days at ambient room temperature, i.e. $T=(22\pm 2)^{\circ}$C. A beam of wavelength $\lambda=1.5418 \AA$ and energy 8~keV is produced by a microsource generator (Rigaku MicroMax 007HF) with a 1200~W rotating anode coupled to a confocal Max-Flux Osmic mirror (Applied Rigaku Technologies, Austin, USA). The dried gel is left for 1~hour in the beam path and diffraction patterns are recorded through a MAR345 image plate detector (MARResearch, Norderstedt, Germany) placed at 152~mm from the sample. The numerical aperture of the detector allows us to probe wavenumbers $q$ from $0.7$~nm$^{-1}$ up to $30$~nm$^{-1}$. Raw data are processed with the software $\copyright$Fit2D: the background intensity diffracted by the surrounding air is first substracted to the raw intensity, which is then normalized by the exposure time and the X-ray beam intensity.

\begin{figure}[!b]
\centering
	\includegraphics[width=0.9\linewidth]{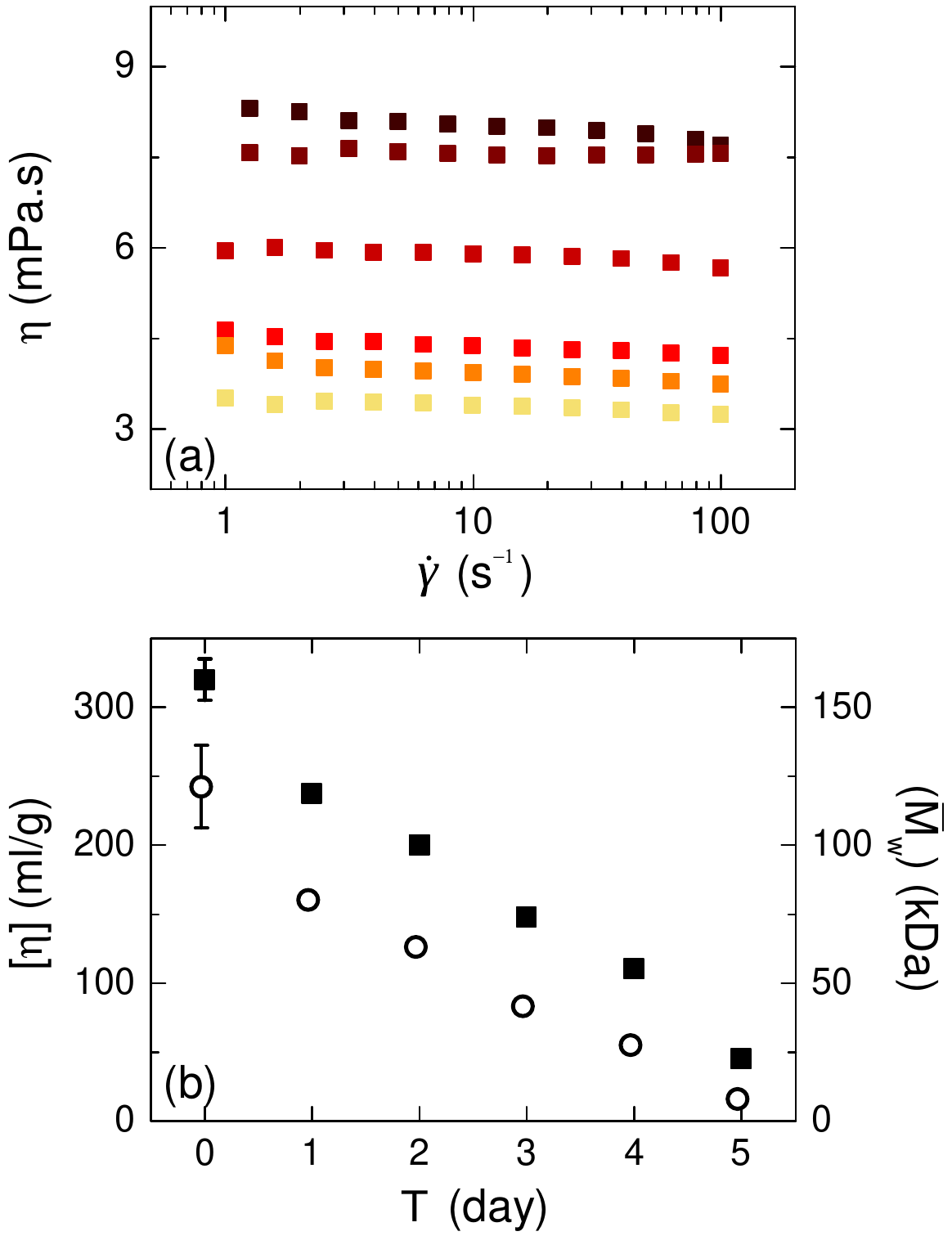}
\caption{(a) Shear viscosity $\eta$ of a 1.5\% w/w agar solution vs. applied shear rate $\dot \gamma$. Colors from black to yellow correspond to different incubation time of the solution at 80$^{\circ}$C: $\mathcal{T}=$1~hour, 1~day, 2, 3, 4 and 5 days. Measurements are performed in a cone and plate geometry at $T=50^{\circ}$C. The shear rate is decreased by step and maintained at each value for 20s. (b) Left: intrinsic viscosity [$\eta$] ($\blacksquare$) as determined by extrapolating viscosity measurements of agar solutions of different concentrations (see text) vs. the incubation time $\mathcal{T}$ of the solution. Error bars result from the average of $\eta$ over the range [1;100]s$^{-1}$ and from the data extrapolation in the limit of vanishing concentrations (see text and Fig.~\ref{fig.s1} in the appendix). Right: average molecular weight $\overline{M}_w$ ($\bigcirc$) computed from the Mark-Houwink formula vs. the incubation time $\mathcal{T}$. The typical error bar is indicated only on the first point for both [$\eta$] and $\overline{M}_w$.  
\label{fig2}}
\end{figure} 

\begin{figure*}[th!]
\centering
	\includegraphics[width=\linewidth]{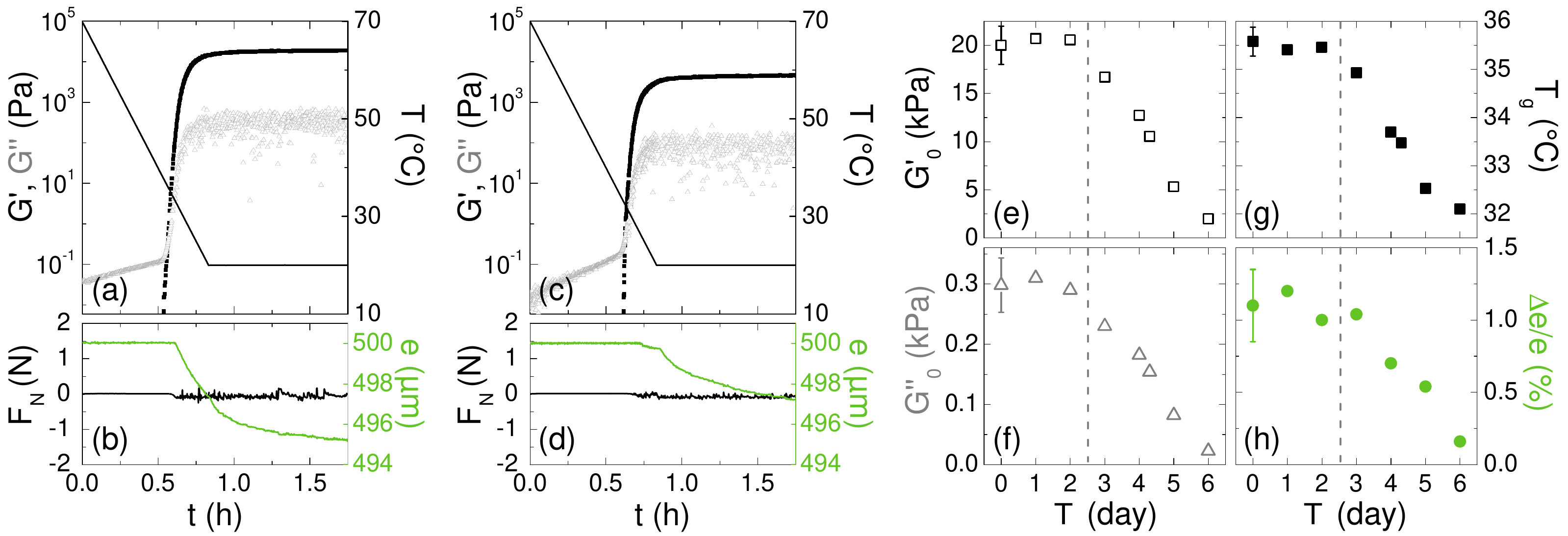}
\caption{(a) Evolution of the elastic ($G'$, $\blacksquare$) and viscous ($G''$, $\triangle$) moduli vs. time $t$ during the cooling of a fresh agar sol ($\mathcal{T} \simeq 1$~hour) from $T=70^{\circ}$C down to $20^{\circ}$C, conducted at $\dot{\rm T}=1^{\circ}$C/min. The crossing of $G'$ and $G''$ defines a gelation temperature $T_g=35.5^{\circ}$C, and for $t\geq 1.75$~h the viscoelastic moduli reach steady state values which are respectively $G'_0=19.2$~kPa and $G''_0=310$~Pa. (b) Imposed normal force $F_N=(0.0\pm0.1)$~N and gap width $e$ vs. time. The gap width, which initial value is $e_0=500~\mu$m, decreases by 1\% due to the sample contraction during the gelation. (c) and (d): same as in (a) and (b) for a gel prepared from an agar sol which has been incubated $\mathcal{T}=5$~days at 80$^{\circ}$C. (e)--(h) Evolution of the visco-elastic moduli $G'_0$ (e) and $G''_0$ (f), the gelation temperature $T_g$ (g) and the relative gap decrease $\Delta e/e$ (h) vs. the incubation times $\mathcal{T}$ of the agar sol prior to the preparation of gels. 
\label{fig3}}
\end{figure*} 

 \section{Heat-induced ageing of the agar sol}
 \label{Solaging}

The present section is devoted to the agar solution, maintained for 5~days at $T=80^{\circ}$C in a testing chamber. The aging of the agar sol is characterized by a series of observations regarding the chemical aging of the polysaccharides. 

 \subsection{Color change and polysaccharides oxidation}
  A fresh solution is prepared in an Erlenmeyer flask as described in section~\ref{Sample}, then sealed and stored in a thermal chamber maintained at 80$^{\circ}$C. The color of the solution turns from transparent color to yellow in a few days, and the color change goes on at least over 12 days [Fig.~\ref{fig.sol}(a)-(d)]. Samples are withdrawn every day during the first 5 days to monitor the evolution of the sol pH, which is measured at 80$^{\circ}$C with a silver electrode (N6000 BNC, Schoot instruments) connected to a controller (C333, Consort). The pH decreases for increasing sample age $\mathcal{T}$ [Fig.~\ref{fig.sol}(e)] which gives evidence of a chemical evolution of the polysaccharide molecules that we attribute to the oxidation of the hydroxyl functional group of the polymers. Indeed, when repeating the experiment on a fresh solution that is stored in contact with an atmosphere continuously enriched in neutral nitrogen instead of ambient air, we observe that the color change and the pH decrease are both strongly reduced [Fig.~\ref{fig.sol}(d)]. The latter experiment proves the key role of oxygen in the ageing process and strongly suggests that the pH decrease can be attributed to the slow oxidation into carboxyl functions of the hydroxyl groups connected to primary carbon atoms in the agar chain [Fig.~\ref{fig0}(Right)].

 \subsection{Viscosity change and chain hydrolysis}

 Concomitantly with the pH measurements, the viscosity of the solution is determined every day through rheological experiments performed in a cone and plate geometry at $T=$50$^{\circ}$C (see section~\ref{Rheol} for technical details). The polysaccharide solution shows an almost Newtonian behavior between 1 and 100~s$^{-1}$ from an hour after being prepared up to $\mathcal{T}=$5~days of incubation at 80$^{\circ}$C [Fig.~\ref{fig2}(a)]. The viscosity of the 1.5\% w/w agar solution decreases for increasing incubation times, which strongly suggests that the polymer chains are becoming shorter due to hydrolysis. To determine the evolution of the molecular weight of the polysaccharides, the same incubation experiment is repeated on agar solutions of different concentrations: $c=0.05$, 0.1 and 0.2\% w/w. The viscosity of each of the three solutions kept all together in the same thermal chamber at $T=80^{\circ}$C is measured every day. A linear extrapolation of the viscosity in the limit of vanishing concentrations allows us to estimate the intrinsic viscosity of the solution defined as $[\eta]= \lim_{c\to 0}\, (\eta-\eta_s)/\eta_sc$, where $\eta_s$ denotes the viscosity of the solvent, here water at $T=50^{\circ}$C [see Fig.~\ref{fig.s1} in the appendix]. The intrinsic viscosity decreases linearly with $\mathcal{T}$ [Fig.~\ref{fig2}(b)]. Moreover, the average molecular weight $\overline{M}_w$ of the polymer chains, which can be estimated from [$\eta$] using the Mark-Houwink formula \cite{Rochas:1989}, goes from 120~kDa for a fresh solution down to 8~kDa after 5 days of incubation at 80$^{\circ}$C which proves the progressive hydrolysis of the polysaccharides.

In conclusion, the color change of the agar sol that goes with an extended heating is due to the oxidation of the primary alcohol functions and the hydrolysis of the polymer chains. The rest of the manuscript is focused on the impact of the sol ageing on the subsequent gels prepared after different incubation times of the agar solution. We will first discuss the evolution of the gelation dynamics and the structural properties of the gel with the incubation time $\mathcal{T}$ of the agar sol, in respectively subsection~\ref{Gelation} and \ref{Gelmic}. Furthermore, the impact of an extended incubation time of the agar sol on both the gel adhesion properties to the metallic plates and the rupture scenario of the gel during macro-indentation tests will be the topic of section~\ref{section5}.
 
\begin{figure*}[!t]
\centering
	\includegraphics[width=\linewidth]{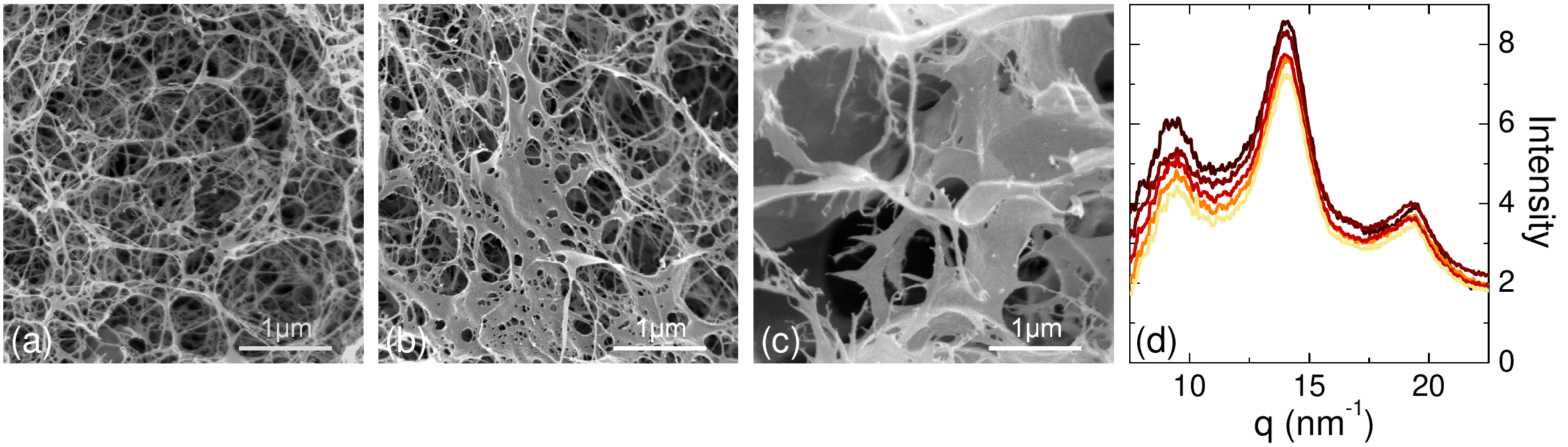}
\caption{(a)--(c) Cryo-SEM images of gels prepared with samples drawn after different incubation times $\mathcal{T}=1$ hour, 3 days and 5 days from a mother solution maintained at $T=80^{\circ}$C. (d) X-ray diffraction spectra $I(q)$, where $q$ stands for the wave number. Colors from black to yellow correspond to gels prepared after different incubation times $\mathcal{T}$ at $T=80^{\circ}$C of the same agar solution: $\mathcal{T}=1$~hour, 1 day, 2, 4 and 5~days.    
\label{fig4}}
\end{figure*} 
 
 \section{Gelation dynamics and gels microstructure}
 \label{Gelcharac}
 
	\subsection{Evolution of the gelation dynamics with $\mathcal{T}$} 
	\label{Gelation}
 Every day, a sample withdrawn from the agar solution maintained at $T=80^{\circ}$C is poured into the pre-heated gap of a parallel-plate geometry (see section~\ref{Rheol} for technical details). The sample is submitted to a decreasing ramp of temperature at a constant rate $\dot{\rm T}=1^{\circ}$C/min down to $20^{\circ}$C to induce gelation, while the elastic and viscous moduli are measured through small amplitude oscillations. Results obtained with a fresh solution ($\mathcal{T}=0$) after only a few hours of incubation of the agar solution at 80$^{\circ}$C are pictured in Fig.~\ref{fig3}(a)--(b). During the first half hour of cooling, $G'$ remains negligible while $G''$ increases continuously. The crossing of $G'$ and $G''$ provides an estimate of the gelation temperature, here refered to as $T_g$. Since the gelation experiment is performed under controlled normal force ($F_N=0.0\pm 0.1$~N) instead of constant gap width, the decrease of the gap width (of about 1\%) compensates for the sample contraction associated with the sol/gel transition \cite{Mao:2015a}. Finally, the elastic and viscous moduli both reach a steady state value after $t=1.5$~h, which are respectively labeled $G'_0$ and $G''_0$. 

The same experiment is repeated every day and the gelation of the agar sol after $\mathcal{T}=5$~days of incubation is reported in Fig.~\ref{fig3}(c)--(d). The gelation occurs later, at a lower temperature ($T_g=$32.5$^{\circ}$C) compared with the case of a fresh agar sol ($T_g=$35.5$^{\circ}$C for $\mathcal{T}\simeq 1$~hour). Moreover, the terminal values of the elastic and viscous modulus of the gel are significantly smaller and the sample is observed to contract only by $\Delta e/e=0.6$\% during the gel formation, compared to the 1\% contraction of the sample prepared from a fresh solution. To go one step further, the systematic evolution of $G'_0$, $G''_0$, $T_g$ and $\Delta e/e$ with the incubation time $\mathcal{T}$ are reported in figures~\ref{fig3}(e)--(h). Gels prepared from a solution younger than 3 days show the same gelation temperature and the same steady-state elastic constants within error bars. However, for $\mathcal{T} \geq 3$~days, the gelation occurs later, i.e. at a lower gelation temperature, and the gel becomes weaker as $\mathcal{T}$ increases: after 5 days of incubation at $80^{\circ}$C, the agar sol leads to a gel that is 75\% weaker than a gel prepared from a fresh solution, which is remarkable. 
We shall emphasize that, although the sol ageing starts from the very moment the sol is stored at 80$^{\circ}$C in the thermal chamber as observed after one day of incubation in Fig.~\ref{fig2}(b), the gelation dynamics and the gel mechanical properties undergo no detectable change before the third day of incubation.

	\subsection{Evolution of the gel microstructure with $\mathcal{T}$} 
	\label{Gelmic}
\begin{figure}[!t]
\centering
	\includegraphics[width=0.9\linewidth]{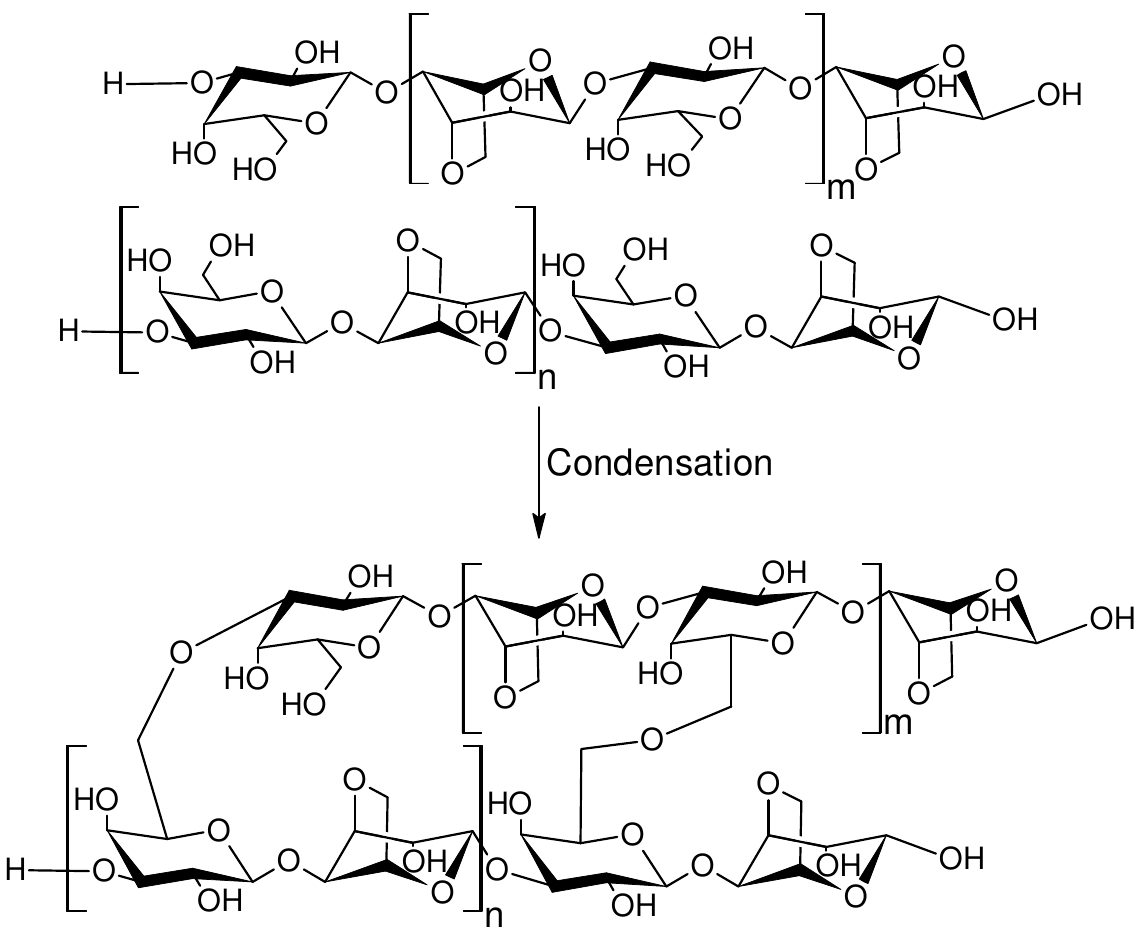}
\caption{Illustration of the intermolecular condensation of the polysaccharides taking place for intermediate and large incubation times of the agar solution at 80$^{\circ}$C. The latter condensation leads to the formation of a foil-like microstructure in gels pictured in Fig.~\ref{fig4}(b) and (c).       
\label{fig4b}}
\end{figure}

 To provide supplemental structural information in addition to the evolution of the bulk mechanical properties, we have performed systematic cryo-SEM observations of gels prepared from samples extracted after different incubation time $\mathcal{T}$ of the solution (see section~\ref{SEM} for technical details). The results are pictured in figure~\ref{fig4}(a)--(c). First, gels formed from a freshly prepared solution show a fibrous-like microstructure composed of interconnected strands delimiting pores with a broad size distribution up to a few micrometers, in agreement with cryo-SEM data reported in the literature \cite{Charlionet:1996,Rahbani:2013}. 
Second, gels prepared after intermediate incubation times of the solution show a coarser microstructure: for $\mathcal{T}=3$~days, gels display a network that is structurally similar to that of a gel prepared from a fresh solution, except for thicker fibers and the formation of some foils [Fig.~\ref{fig4}(b)]. Last, gels made after an incubation time $\mathcal{T}=5$~days of the solution only show a few fibers of comparable size to those of the gel prepared from a fresh solution. Most of the gel structure now consists in foils larger than a few microns [Fig.~\ref{fig4}(c)]. We attribute such evolution of the gel structure for increasing incubation times of the solution, to the condensation of the polysaccharides in solution as the heated solution slightly scatters the incident light after 5 days incubation [Fig.~\ref{fig4b}]. Indeed, the intermolecular condensation between polysaccharides is favored by the acidification of the solution that goes with increasing incubation times [Fig.~\ref{fig.sol}(e)] and accounts for the growth of a large scale microstructure.   

\begin{figure}[!t]
\centering
	\includegraphics[width=0.9\linewidth]{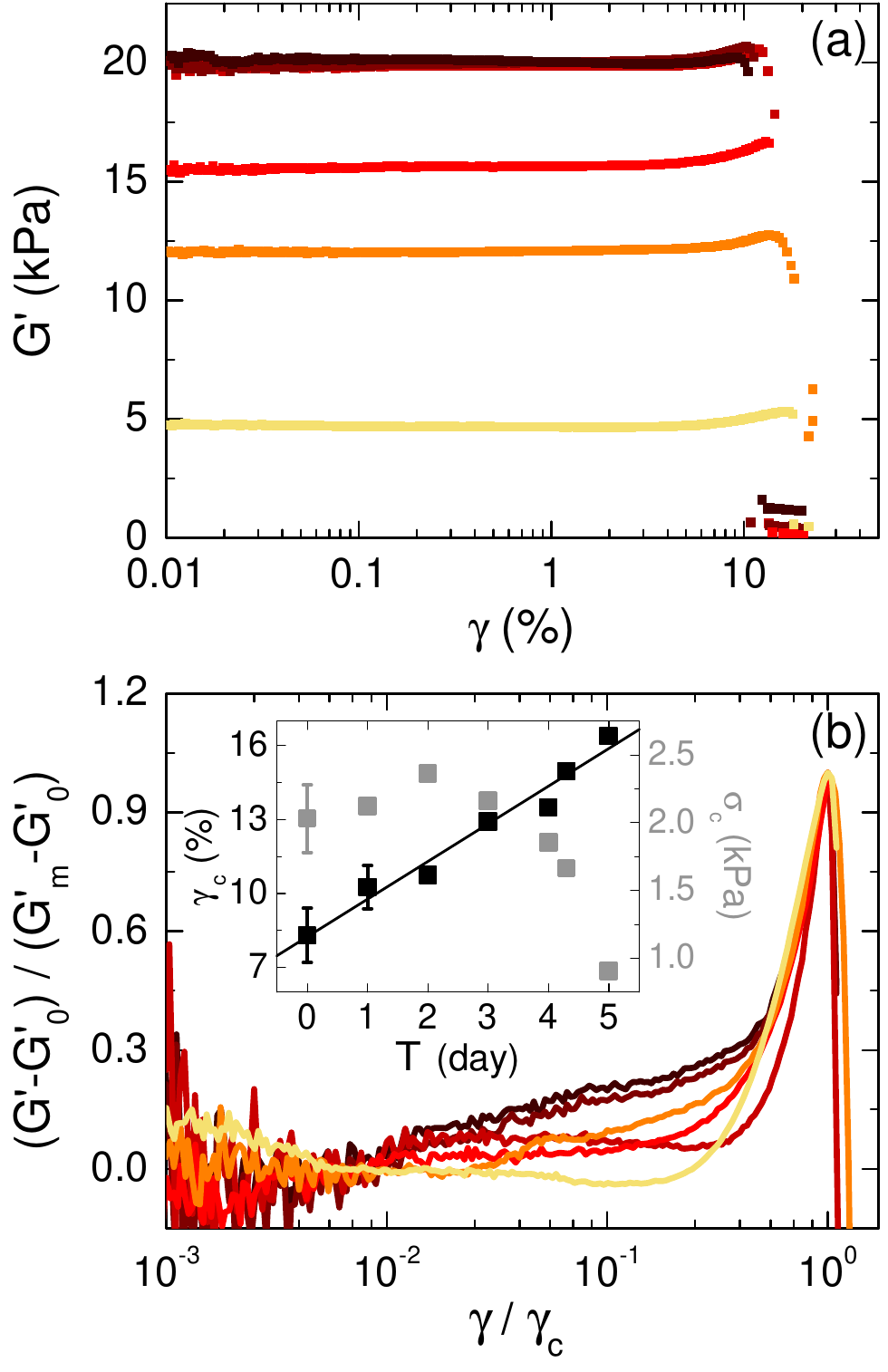}
\caption{(a) Elastic modulus $G'$ vs. strain $\gamma$ monitored during a strain sweep experiment performed at $f=1$~Hz. The strain increases logarithmically from $\gamma=0.01$\% to 100\% over a duration of 2160s (9s/point). Colors from black to yellow correspond to gels obtained after different incubation times of the same agar solution at 80$^{\circ}$C: $\mathcal{T}=$1~hour, 1~day, 2, 3, 4 and 5 days. (b) Normalized elastic modulus $(G'-G'_0)/(G'_{\rm m}-G'_0)$ where $G'_0=G'(\gamma=0.1\%)$ vs. normalized strain $\gamma/\gamma_c$, where $\gamma_c$ denotes the critical strain at which $G'$ is maximum. Inset: Evolution of $\gamma_c$ ($\blacksquare$) and $\sigma_c=\sigma(\gamma_c)$ (\textcolor{gray}{$\blacksquare$}) vs. the incubation time $\mathcal{T}$ of the agar sol. The black line corresponds to the best linear fit of the data: $\gamma_c=7.8+1.7\mathcal{T}$. Error bars were determined by repeating the experiments on three different samples.       
\label{fig5}}
\end{figure} 

To probe smaller scales, and determine whether or not the fibrous and foil-like microstructures are made from the same elementary components, we perform X-ray diffraction experiments on gels prepared after different incubation times $\mathcal{T}$ of the agar sol. Experiments are performed on dried samples to avoid the diffraction spectra associated with water (see section~\ref{Xray} for technical details and Fig.~\ref{fig.s2} in the appendix). Results are reported in Fig.~\ref{fig4}(d). The spectra associated with a gel obtained from a fresh solution ($\mathcal{T}$ of a few hours) shows 3 maxima at the following wavenumbers: $q_1=9.45~\mathrm{nm}^{-1}$, $q_2=13.86~\mathrm{nm}^{-1}$ and $q_3=19.3~\mathrm{nm}^{-1}$ which correspond respectively to: $d_1 \equiv 2\pi/q_1=0.66$~nm, $d_2=0.45$~nm and $d_3=0.32$~nm, in quantitative agreement with the seminal work of Foord and Atkins \cite{Foord:1989}. The agarose chains are associated into 3-fold double helices, and $q_1$ and $q_3$ correspond respectively to the distance between two different and two identical saccharides, while $q_2$ stands for the helix diameter. For increasing incubation times $\mathcal{T}$, the shape of the diffraction spectra and the positions of the maxima remain identical, which proves that the building blocs of the gel remain the same despite the slow heat-induced ageing of the agar sol. The decay of the maxima amplitude for increasing incubation times $\mathcal{T}$ further indicates that an increasing fraction of polymer chains is not involved in double helix formation during the gelation process, which is compatible with an increasing amount of intermolecular condensation.      

\begin{figure*}[!t]
\centering
	\includegraphics[width=\linewidth]{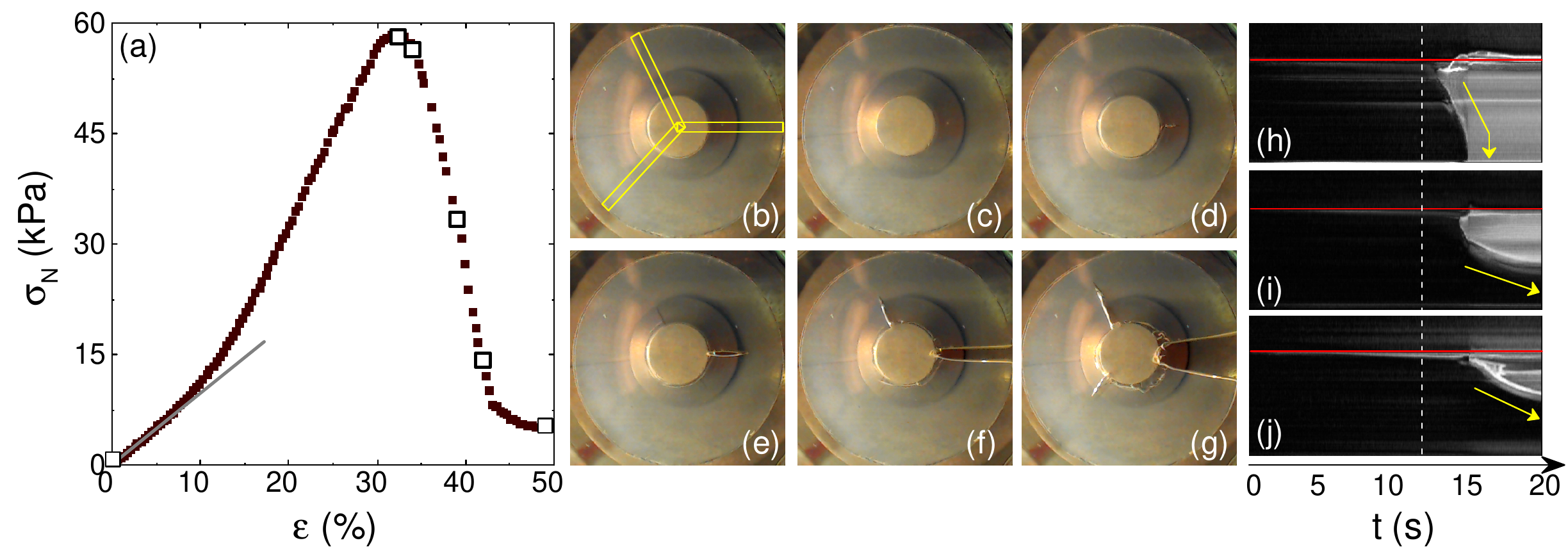}
\caption{(a) Normal stress $\sigma_N$ vs. strain $\epsilon$ during the indentation experiment with a flat-ended cylinder, of a gel disk prepared from a fresh solution ($\mathcal{T}$ of a few hours). (b)--(g) Images taken during the indentation of the disk at $\epsilon=0$\%, 32.7\% (stress maximum), 34.2\% (nucleation of the first macroscopic crack), 39.1\%, 41.9\% (the first crack reaches the outer edge of the agar disk) and 48\%. These deformations are emphasized by open squares in (a). The spatial scale is set by the disk diameter of 34~mm. (h)--(j) Spatio-temporal diagrams obtained from the orthoradial projection of the grey levels within the three radial sections that encompass the fractures [yellow rectangles in (b)]. For all the diagrams, the first image is substracted from the stack of images to improve the contrast and better visualize the crack propagation. The upper and lower limits of each diagram correspond respectively to the center and the peripheral region of the gel disk, so that the height of each diagram represents the gel disk radius of 17~mm. The horizontal red line refers to the outer region of the indenter. The white vertical dashed line marks the time for which the stress reaches the maximum value observed on the mechanical response reported in (a). The slope of the yellow arrows in (h)--(j) gives an estimate of the crack front velocity.          
\label{fig7}}
\end{figure*} 

\section{Gel debonding and failure dynamics}
 \label{section5}
 This last section is dedicated to the impact of the incubation time $\mathcal{T}$ of the agar sol on the non-linear rheological properties and the yielding scenario of the agar gels. First, we quantify the evolution of the gel adhesion properties to the metallic plates by performing strain sweep experiments. Second, we report the spatially resolved scenario of the gel failure through benchmark macro-indentation experiments, and we discuss the evolution of the latter scenario with $\mathcal{T}$.

 	\subsection{Gel debonding from the plates}\label{strainsweep}
The gelation experiments which are reported in Fig.~\ref{fig3} and conducted in a parallel-plate geometry under controlled zero normal force, end after two hours when both the gel elastic modulus $G'$ and the gel thickness $e$ reach steady state values. The rheometer is then switched from controlled normal force to constant gap width, maintaining the latest value of $e$, after which a strain sweep experiment is performed to quantify the gel adhesion to the metallic geometry. Strain controlled oscillations of frequency 1~Hz and increasing amplitude from $\gamma=$0.01\% up to 100\% are imposed to the gel over a total duration of 2160~s. Gels prepared after different incubation times display a similar response pictured in Fig.~\ref{fig5}(a): the elastic modulus remains constant at low strain values, whereas it increases at larger strain amplitudes to reach a maximum value $G'_{\rm m}$ at $\gamma=\gamma_c$ which depends on the incubation time. Finally, $G'$ decreases sharply above $\gamma_c$. The increase of $G'$ under external stress corresponds to the strain hardening of the agar gel, classically observed for biopolymer gels \cite{Groot:1996,Storm:2005,Pouzot:2006,Zhang:2007,Brenner:2009,Carrillo:2013} including agar(ose) gels \cite{Nakauma:2014}. The subsequent and abrupt decay of $G'$ for strains larger than $\gamma_c$ can be interpreted as the sudden debonding of the gel from the upper plate. The latter conclusion is supported by the fact that the gel remains intact and shows no macroscopic crack when raising the upper plate after the strain sweep test.  
Therefore $\gamma_c$ provides an estimate of the critical deformation above which the gel detaches from the metallic walls of the shear cell as a result of interfacial debonding and water release, although we cannot rule out the presence of microscopic failure inside the gel. Furthermore, the strain responses $G'(\gamma)$ of gels prepared for increasing incubation times can be superimposed by plotting $(G'-G'_0)/(G'_{\rm m}-G'_0)$ vs. $\gamma/\gamma_c$, where $G'_0$ denotes the steady-state elastic modulus determined in the linear regime [Fig.~\ref{fig5}(b)]. Such a rescaling suggests that the same debonding scenario is at stake for all the strain sweep experiments, and since the critical strain $\gamma_c$ increases linearly with the incubation time $\mathcal{T}$ [inset in Fig.~\ref{fig5}(b)], one can conclude that gels prepared from sol heated over longer durations stick better to the plates and release water less easily under shear. The ability of weaker gels to dissipate energy and undergo plastic deformations leads for increasing $\mathcal{T}$ to larger deformations before debonding. In other words, the transition from elastic friction to hydrodynamic lubrication under shear occurs at a larger critical strain for gels prepared after prolonged incubation of the agar solution. 
Nonetheless, we shall emphasize that $\gamma_c$ is characteristic of the boundary conditions, and that the use of other materials for the upper and lower plates would lead to different values and/or dependence of $\gamma_c$ with $\mathcal{T}$, which extensive study is out of scope of the present work. 
Finally, we note that contrary to the critical strain $\gamma_c$, the critical stress defined as $\sigma_c=\sigma(\gamma_c)$ bears the signature of the aging of the agar sol, as do the gel elastic moduli [Fig.~\ref{fig3}(e)-(f)]. Indeed, for incubation time shorter than 3~days, the stress at debonding $\sigma_c$ is constant independent of $\mathcal{T}$, whereas it decreases for gels prepared from a solution incubated for more than 3~days [inset Fig.~\ref{fig5}(b)].

	\subsection{Macro-indentation experiments} \label{macroindent}
To characterize the bulk mechanical properties of the gel in the non-linear regime, macro-indentation tests are performed with a flat-ended cylinder on agar gels disks that are prepared after different incubation times $\mathcal{T}$ of the same solution. The deformation and the rupture of the gel are monitored from below through a transparent bottom plate by means of a high resolution webcam, while the normal force sensor of the rheometer is used to record the stress-strain curve $\sigma(\epsilon)$ (see section~\ref{Rheol} for technical details). 
 
Results for a gel prepared from a fresh solution ($\mathcal{T} \simeq 1$~h) are reported in Fig.~\ref{fig7} and in the supplemental movie~1 in the appendix. The stress increases linearly with the strain up to a few percents, followed by a faster than linear increase associated with the strain hardening of the gel, up to a maximum value $\sigma_N^{(m)}$ at $\epsilon=\epsilon^{(m)}$. Meanwhile, the gel does not show any damage visible by the naked eye neither in the linear regime, nor in the strain hardening regime [Fig.~\ref{fig7}(b)] and remains intact up to the stress maximum [Fig.~\ref{fig7}(c)]. The first fracture becomes visible right after the stress maximum, and nucleates at a finite distance from the indenter [Fig.~\ref{fig7}(d)]. As the strain increases beyond $\epsilon^{(m)}$, the stress rapidly decays towards a low plateau value, while two other fractures nucleate at an angle of about 120${^\circ}$ of the first fracture [Fig.~\ref{fig7}(e) and (f)]. The three fractures grow along the disk radius and propagate towards the edge of the gel [Fig.~\ref{fig7}(g)].
 A spatio-temporal plot consisting in an orthoradial projection of the grey levels within a thin radial section of the gel disk pictured in yellow in Fig.~\ref{fig7}(b), confirms that the first fracture indeed nucleates at a finite distance from the indenter [Fig.~\ref{fig7}(h)]. Furthermore, the former representation shows that the tip of the first fracture grows at a constant speed $u_1=(4.7\pm0.4)$~mm/s up to about 3~mm from the edge of the disk, where the fracture opening suddenly accelerates. The second and third fractures grow at lower velocities than $u_1$, which both decrease as the fractures propagate towards the edge of the disk [Fig.~\ref{fig7}(i) and (j)]. 
 
The linear low-strain regime allows us to define an apparent compression modulus $E^*=\partial\sigma/\partial\epsilon$ as the initial slope of the stress-strain curve. Taking into account the cylindrical shape of the indenter, the finite thickness $H$ of the gel disk and the stress singularity at the sharp edge of the indenter, the true compression modulus $E$ of the material obeys the following relation derived from previous analysis \cite{Hayes:1972,Haider:1997}:
\begin{equation}
E=(1-\nu^2)\frac{r}{H}\frac{\pi}{2\kappa}E^*
\label{Modulus}
\end{equation}
where $\nu$ denotes the Poisson coefficient of the gel, $r=5$~mm is the radius of the indenter, $H=3.9$~mm stands for the thickness of the gel disk and $\kappa$ is a numerical correction related to the finite size of the system which depends on $r$, $H$ and $\nu$. The only unknown is the Poisson coefficient $\nu$ that is otherwise carefully determined through indentation experiments performed on gel disks of different thicknesses [see Fig.~\ref{fig.s3} and the corresponding discussion in the appendix]. We find $0.1\leq \nu \leq 0.3$ which leads to a coefficient $\kappa$ such as $2.6\leq \kappa \leq 3$ using the expression provided in ref. \cite{Haider:1997} and therefore to $E=(60\pm 10)$~kPa for a gel prepared from a fresh agar sol. Note that the latter value is in quantitative agreement with the shear modulus $G'=(20\pm 2)$~kPa measured through small amplitude oscillations [Fig.~\ref{fig3}] since $2G'(1+\nu)=(48\pm 10)$~kPa is compatible with $E$ within error bars, showing that the agar disk behaves as an isotropic material \cite{Landau:1970}. 

\begin{figure}[t!]
\centering
	\includegraphics[width=\linewidth]{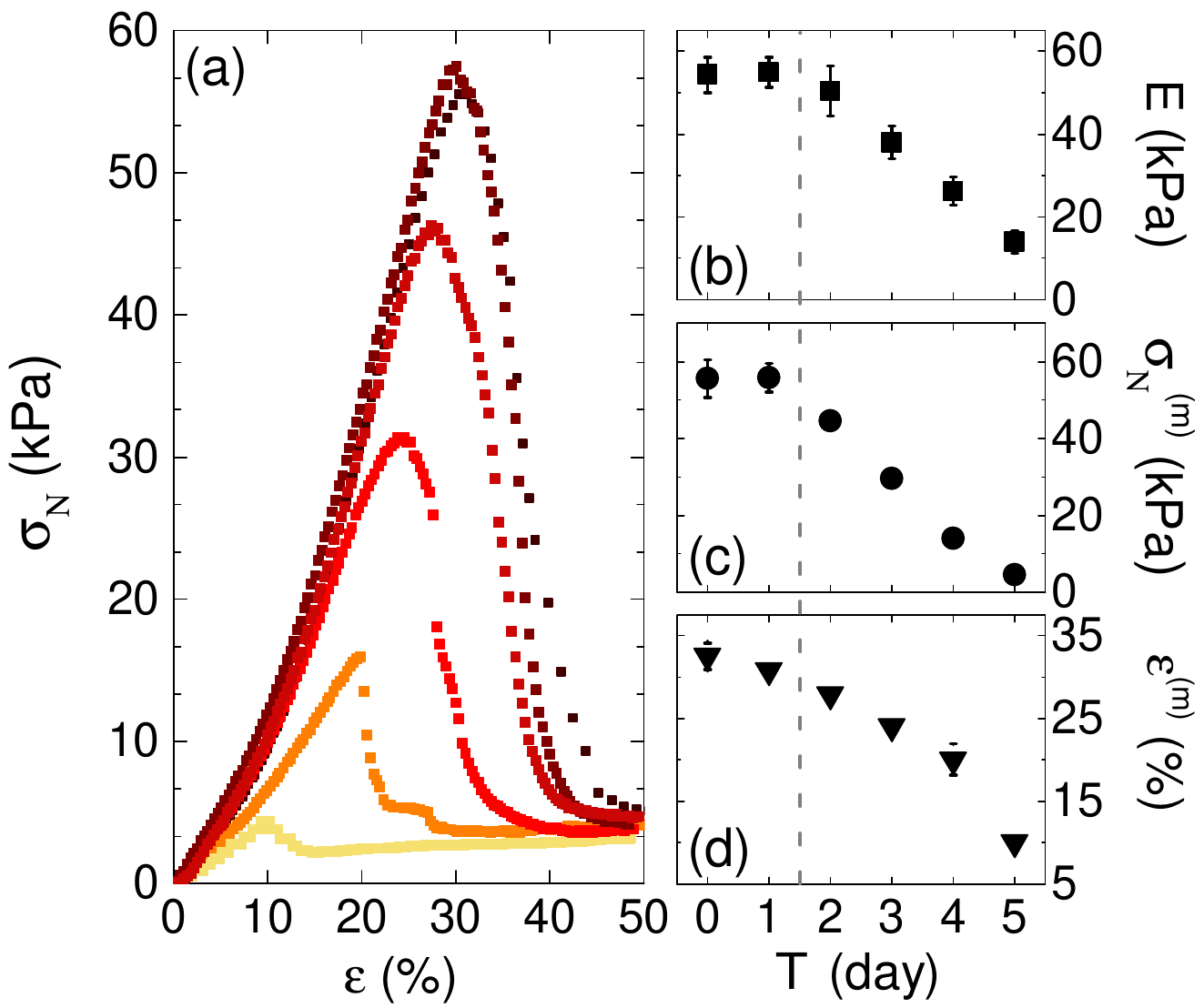}
\caption{(a) Normal stress $\sigma_N$ vs. strain $\epsilon$ monitored during macro-indentation experiments. Colors from black to yellow stand for gels prepared from the same solution which is left at $T=80^{\circ}$C for different incubation times: $\mathcal{T}=1$~hour, 1 day, 2, 3, 4 and 5 days. (b) Compression modulus $E$ of the gel vs. the incubation time $\mathcal{T}$ of the agar sol when considering a Poisson coefficient of $\nu =0.3$. (c) Stress maximum $\sigma_N^{(m)}$ reached during the indentation vs. $\mathcal{T}$. (d) Strain $\epsilon^{(m)}$ associated with the stress maximum vs. $\mathcal{T}$. Error bars were determined by repeating the experiments on three different samples.       
\label{fig6}}
\end{figure} 

\begin{figure*}[t!]
\centering
\includegraphics[width=\linewidth]{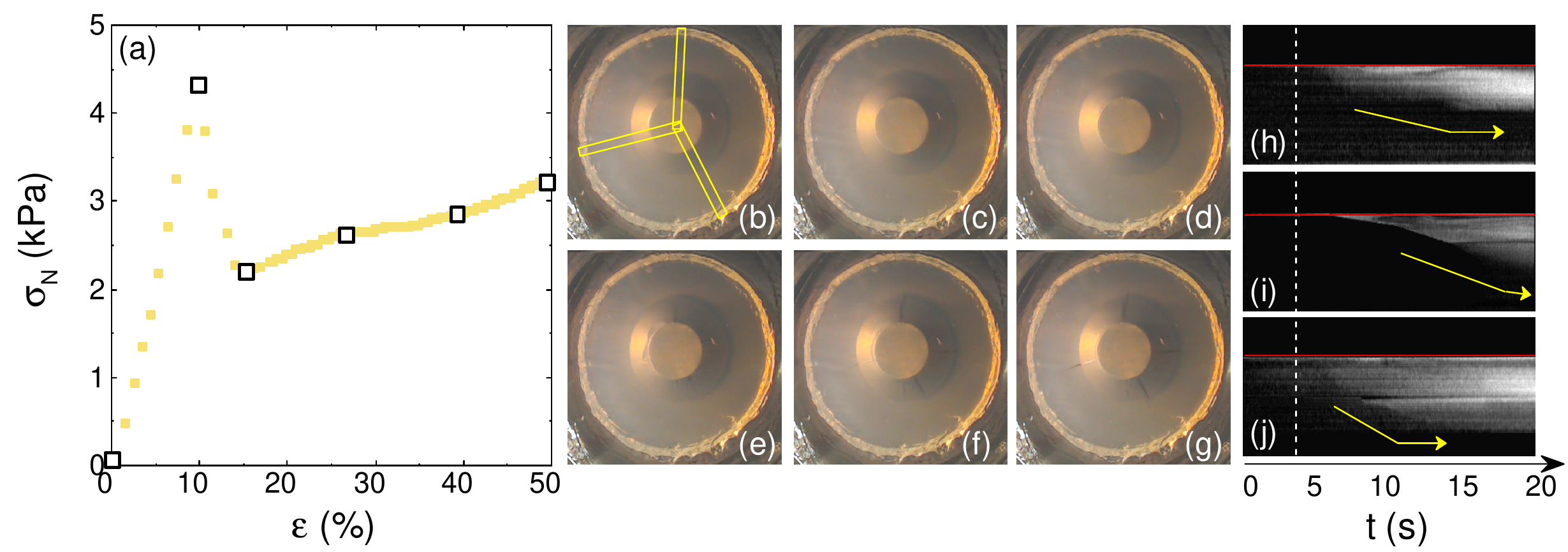}
\caption{ (Color online) Normal stress $\sigma_N$ vs. strain $\epsilon$ during a macro-indentation experiment of a gel disk made from a solution incubated during $\mathcal{T}=5$~days. (b)--(g) Images taken during the indentation of the disk at respectively $\epsilon=0$, $\epsilon=$9.8\% (stress maximum), $\epsilon=15.1$\%, 26.2\%, 39.1\% and 48.6\%. These deformations are emphasized by open squares in (a). The spatial scale is set by the disk diameter of 34~mm. (h)--(j) Spatio-temporal diagrams obtained from the orthoradial projection of the grey levels within the three radial sections that encompass the first three fractures that grow during the test [yellow rectangles in (b)]. For all the diagrams, the first image is subtracted from the stack of images to improve the contrast and better visualize the crack propagation. The upper and lower limits of each diagram correspond respectively to the center and the peripheral region of the gel disk, so that the height of each diagram represents the gel disk radius of 17~mm. The horizontal red line refers to the outer edge of the indenter. The white vertical dashed line marks the time at which the stress reaches the maximum value observed on the mechanical response reported in (a). The slope of the yellow arrows in (h)-(j) gives estimate of the crack front velocity.
\label{fig10}}
\end{figure*} 

Building upon this first test, the exact same indentation experiment is repeated on agar gels prepared after different incubation times $\mathcal{T}$ of the agar solution. Data are reported in Fig.~\ref{fig6}(a). All the gels display a stress response of similar shape to the one discussed above for $\mathcal{T}=0$: the stress first increases linearly for small strain values ($\epsilon\leq 5$\%), then the sample shows some strain hardening before reaching a rupture point which coordinates ($\epsilon^{(m)}$, $\sigma_N^{(m)}$) strongly depends on the incubation time. Beyond the stress maximum, a first fracture nucleates in the vicinity of the indenter, and the stress shows a rapid decay towards a plateau value concomitantly to the nucleation and growth from the intender of two or three supplemental radial fractures. Systematic evolution of $E$ as well as $\sigma_N^{(m)}$ and $\epsilon^{(m)}$ with the incubation time $\mathcal{T}$ of the agar sol are reported in Fig.~\ref{fig6}(b)--(d). For $\mathcal{T}\leq 2$~days, the three parameters show consistent values within error bars, illustrating that the compression modulus and the rupture point remain the same despite some chemical degradation of the polysaccharides has already started (Fig.~\ref{fig2}). For $\mathcal{T}\geq 3$~days, the compression modulus decreases linearly to reach one tenth of its initial value for a gel prepared from a sol incubated for $\mathcal{T}=5$~days. Furthermore, gels fail at smaller strain values for increasing incubation times $\mathcal{T}$: from $\epsilon^{(m)}\simeq$30\% for a gel prepared after $\mathcal{T}=1$~day of incubation of the agar sol, down to $\epsilon^{(m)}\simeq$10\% for $\mathcal{T}=5$~days [Fig.~\ref{fig10}(a)]. 
The failure scenario of the latter gel shows quantitative differences with that reported for a fresh gel in Fig.~\ref{fig7}. Indeed, for a gel prepared after $\mathcal{T}=5$~days of incubation of the agar sol, the stress maximum is associated with the formation of a plastic region in the vicinity of the indenter [see the left side of the indenter in Fig.~\ref{fig10}(d)], and fractures only grow at strain values much larger $\epsilon^{(m)}$ [Fig.~\ref{fig10}(e)-(g) and see also movie~2 in the appendix]. 

\begin{figure}[!t]
\centering
	\includegraphics[width=\linewidth]{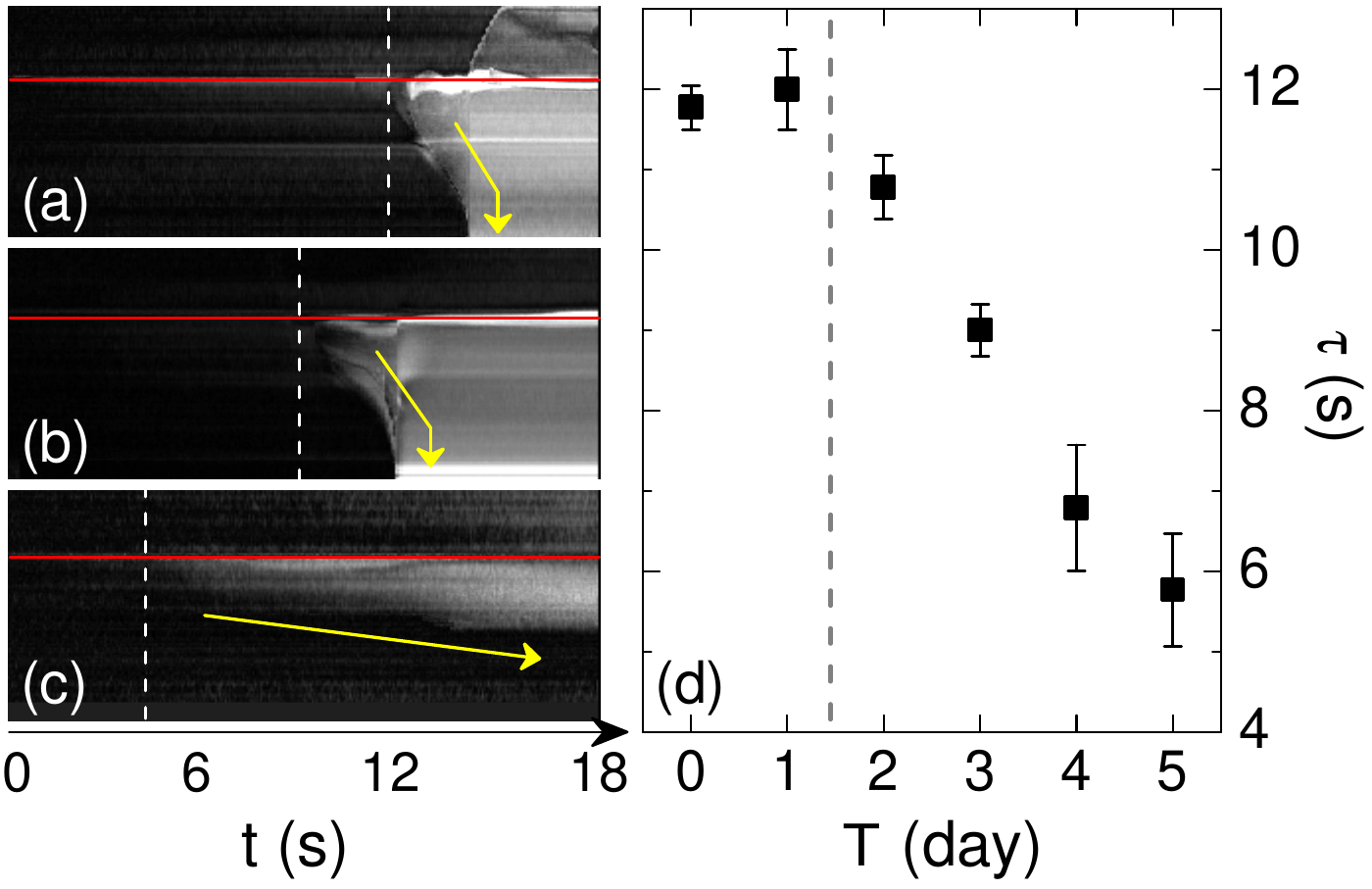}
\caption{(a)-(c) Spatio-temporal diagrams of the radial section of the gel disk that encompass the first fracture, for gels prepared at different incubation times. From top to bottom: $\mathcal{T}=1$, 3 and 5~days. For each diagram, the first image is substracted from the stack of images for a better contrast. The horizontal red line refers to the outer edge of the indenter while the height of each diagram represents the gel disk radius of 17mm. The white vertical dashed lines mark the time at which the  stress reaches the maximum observed in Fig.~\ref{fig6}(a). The slope of the yellow arrows in (a)-(c) illustrate the velocity of the crack front. (d) Nucleation time $\tau$ of the first fracture observed during macro-indentation experiments of gels obtained after different incubation time $\mathcal{T}$ of the same agar solution at $T=80^{\circ}$C.       
\label{fig8}}
\end{figure}

As a characteristic observable of the difference in the rupture process of gels prepared after different incubation times $\mathcal{T}$, we consider the nucleation time $\tau$ of the first fracture, as visible by eye. The first fracture appears sooner for increasing values of the incubation time, as clearly observed in the spatio-temporal diagrams of the gel radial section which encompass the first fracture for gels prepared after $\mathcal{T}=$1, 3 and 5 days [Fig.~\ref{fig8}(a)--(c)]. More quantitatively, the nucleation time of the first fracture starts decreasing for $\mathcal{T}\geq 2$~days [Fig.~\ref{fig8}(d)], in a similar fashion to the behavior of the strain $\epsilon^{(m)}$ associated with the stress maximum, which also decreases after two days of incubation, as discussed above [Fig.~\ref{fig6}(d)]. Furthermore, although the first fracture appears sooner for increasing incubation times, fractures become less visible and propagate more slowly as illustrated by the slope of the yellow arrows in Fig.~\ref{fig8}(a)-(c). Finally, the crack propagation becomes much more localized in the radial direction, for gels prepared from a solution incubated for more than two days  [Fig.~\ref{fig10}(b)-(g) and see also movie~2 in the appendix]. Indeed, for increasing incubation times $\mathcal{T}$, agar gels exhibit a more ductile behavior compared to the brittle-like rupture scenario reported in Figure~\ref{fig7} for gels prepared from a fresh agar sol. Such a brittle to ductile transition is clearly visible in the post-mortem images of the gel disks [compare Fig.~\ref{fig7}(g) to Fig.~\ref{fig10}(g)]. In brief, a prolonged incubation of the agar solution produces weaker gel through which fractures propagate more slowly because of the gel ductile behavior. The latter behavior is also responsible for the increase of the critical strain $\gamma_c$ above which the gel detaches from the plates under shear, as reported above in Fig.~\ref{fig5}(b).

\section{Discussion}

We can conclude from section~\ref{Solaging} that the agarose molecules of an agar solution stored at $T=80^{\circ}$C experience hydrolysis and intramolecular oxidation less than a day after starting the incubation. Yet, the impact on the gelation of the solution ageing is only visible after three days of incubation, at pH$\simeq 5$. Indeed, the intermolecular oxidation of the agarose molecules, which is responsible for the change in structural and mechanical properties of the agar gels only occurs above three days of incubation. We shall emphasize that the exact value of three days depends on the temperature at which the agar solution is stored, and also on the quality of the agar that is used which will affect the evolution of the pH. Indeed, we have repeated the same test as the one reported in Fig.~\ref{fig3} on a different agar brand (BioM\'erieux instead of Sigma-Aldrich). In the latter case, the elastic shear modulus starts decreasing for incubation times longer than five days [See Fig.~\ref{fig.s5} in the appendix] instead of three [Fig.~\ref{fig3}]. Therefore, any estimate of the critical duration $\mathcal{T}_c$ below which the incubation period of an agar sol has no effect on the corresponding agar gels should be determined anew for every agar-based sample of interest, especially if that sample contains additional components which may impact the pH and thus the oxidation and the hydrolysis of the agarose.  

 For $\mathcal{T}>\mathcal{T}_c$, the agar gel microstructure becomes coarser as the result of intermolecular condensation of the agarose molecules. The growth of micron-sized foils during the gelation, give rise to larger pores in the microstructure of the gels leading to both weaker shear and compression modulus. A few days of incubation of the agar sol allow us to lower the elastic properties by 80\% which illustrates how the incubation could be used to modify the gel mechanical properties. Concomitantly to the decrease of the elastic properties, gels made from overcooked agar sol exhibit smaller yield strains in compression and a more ductile behavior as fractures created by macro-indentation remain localized close from the indenter. This ductile behavior also results in an increase of the critical strain above which the gel debonds from the plate-plate geometry during oscillatory shear experiments of increasing strain amplitude.
Such a change of both the structure and the mechanical properties is strongly reminiscent of that observed in gels prepared from a fresh solution and loaded with sucrose \cite{Deszczynski:2003}. Indeed the sucrose is easily oxidized at high temperature, quickly leading to a pH decrease which in turn favors the intermolecular condensation responsible for the formation of micron-sized foils.
 
 Finally, aside of macro-indentation experiments and oscillatory shear tests, we have shown that a 1.5\% w/w agar gel prepared from a fresh agar solution exhibits a Poisson coefficient such as $0.1 \leq \nu \leq 0.3$. That value is quantitatively smaller than the value of 0.5 commonly admitted in the literature \cite{Chen:2001,Scandiucci:2006,Norziah:2006} and shows that 1.5\% w/w agar gels are somewhat compressible. Experimentally, the main uncertainty on $\nu$ comes from the determination of $G'$ which is extremely delicate. Only the zero normal force protocol that we use here allows to compensate for the gel contraction that otherwise strongly affects the values $G'$ determined with a constant gap geometry \cite{Mao:2015a}. Indeed, the shear elastic modulus reported in ref. \cite{Norziah:2006} and measured at constant gap, shows either a drift at long times associated with strain hardening or abrupt drops during the gelation that both result from the contraction of the sample \cite{Mao:2015a}, and may lead to a systematic error on $G'$. Systematic measurements of Poisson coefficients for agar gels of various shapes and polymer contents are out of the scope of the present publication and will be reported in detail elsewhere.  

\section{Conclusion}	
     
 We have shown that the structural and mechanical properties of agar gels can be tuned via the incubation time of the agar solution at high temperature. Modifications of the gel properties result from the intramolecular oxidation and the hydrolysis of the polysaccharides in solution that lead to a pH decrease, which in turn favors the intermolecular condensation of the polysaccharides during the gelation. Gels prepared from overcooked solutions show an increasingly coarser microstructure associated with softer elastic properties, and a ductile-like behavior that makes the gel less resistant in compression and delays the gel debonding from the plates when undergoing a shear deformation. The present study rationalizes pioneering observations \cite{Whyte:1984} and provides a quantitative estimate for the critical duration below which an agar solution can be stored at high temperature without affecting the properties of the subsequent gels. As such, our result should serve in the future as a guideline for the design of efficient manufacturing processes of agar-based materials.

\section*{Acknowledgments}    
This work received funding from BioM\'erieux and the ANRT under the CIFRE program, Grant Agreement No.~112972. The authors gratefully acknowledge H.~Saadaoui for AFM measurements of the plates surface roughness, P.~Legros for his help with the cryo-SEM experiments, as well as T.~Gibaud and F.~Villeval for stimulating discussions.


\appendix

\begin{figure}[t!]
\centering
\includegraphics[width=0.9\linewidth]{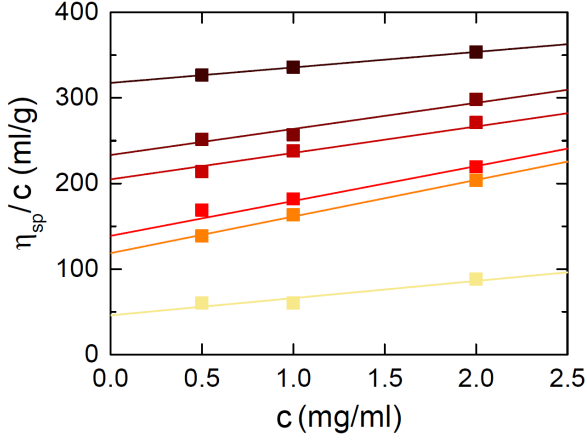}
\caption{ (Color online) Reduced viscosity $(\eta- \eta_s)/\eta_sc$ vs. the polymer concentration $c$, where $\eta_s$ denotes the viscosity of the solvent (water at $T=50^{\circ}$C). Colors from black to yellow correspond to different incubation times of the solutions at $T=80^{\circ}$C: $\mathcal{T}=$1~hour, 1~day, 2 days, 3 days, 4 days and 5 days. Measurements are performed in a cone and plate geometry at $T=50^{\circ}$C. Each point corresponds to an average over the following range of shear rate: $1\leq \dot \gamma \leq 100$~s$^{-1}$. Lines denote the best linear fit of the data and the intercept provides an estimate of the intrinsic viscosity $[\eta]$ of the solution as a function of the incubation time $\mathcal{T}$.   
\label{fig.s1}}
\end{figure} 

\section{Supplemental movies}

\textbf{Supplemental movie~1} shows the failure of an agar gel prepared from a fresh agar sol ($\mathcal{T}\simeq1$~hour) together with the simultaneous evolution of the normal stress during a macro-indentation experiment. The corresponding data are pictured in Fig.~8 in the main text. \textbf{Supplemental movie~2} shows the result for a similar indentation experiment performed on a gel prepared from the same agar sol which has been submitted to $\mathcal{T}=5$~days of incubation at $T=80^{\circ}$C. The corresponding data are pictured in Fig.~10 in the main text.

\section{Supplemental figures} 

\textbf{Supplemental Fig.~\ref{fig.s1}} shows the viscosity of three solutions of different agar concentrations: $c=0.5$, $1$ and $2$~mg/mL, determined after different incubation times $\mathcal{T}$ ranging from a few hours to 5 days (colors from black to yellow). The three solutions are stored together in the same thermal chamber to ensure the same thermal history. The viscosity of samples drawn at regular time intervals are determined by steady shear experiments over the following range of shear rates $1\leq \dot \gamma \leq 100$~s$^{-1}$ (see subsection~2.2.1 in the main text for technical details).

\begin{figure}[t!]
\centering
\includegraphics[width=0.9\linewidth]{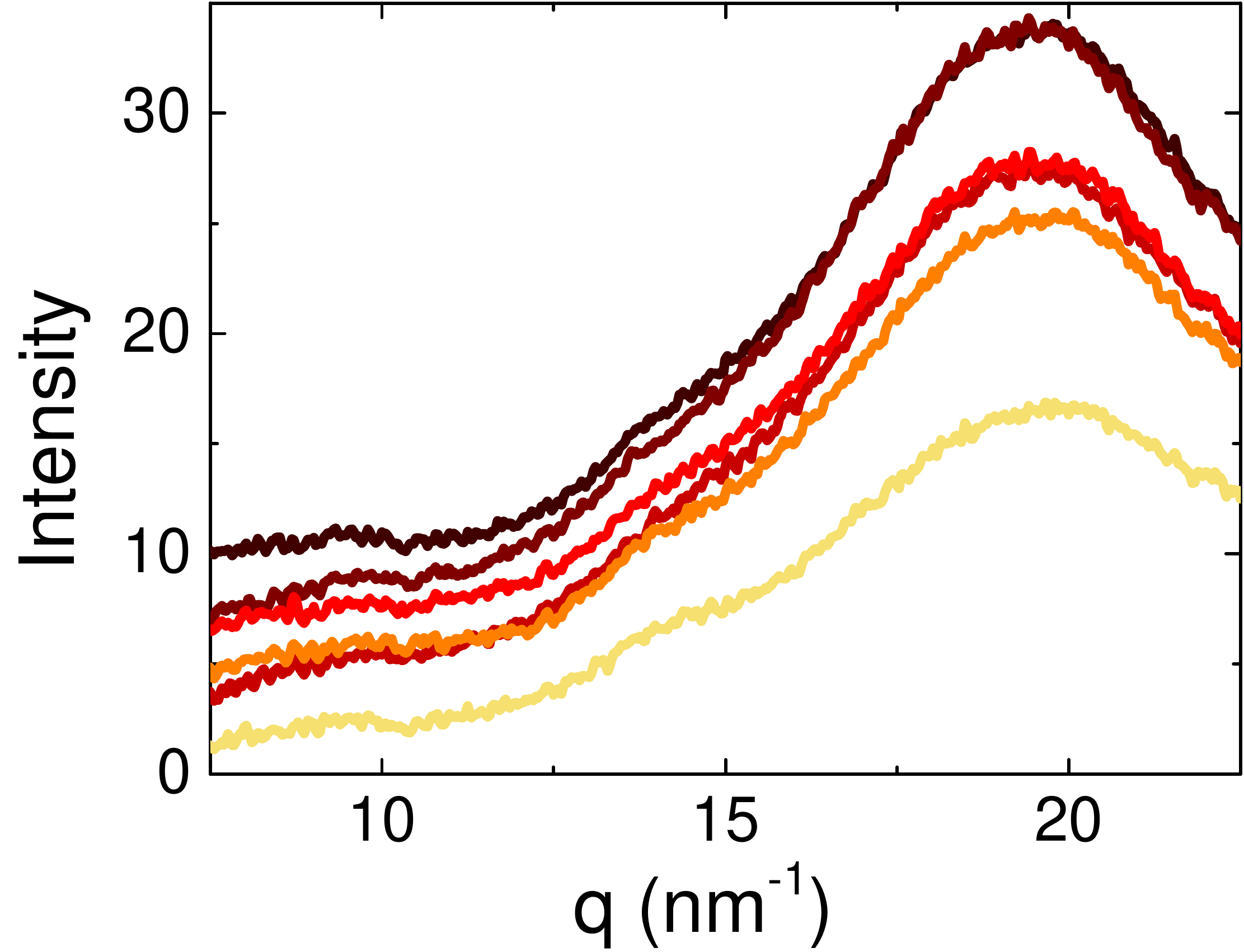}
\caption{ (Color online) X-ray diffraction spectra $I(q)$, where $q$ stands for the wavenumber. Colors from black to yellow correspond to gels prepared after different incubation times at $T=80^{\circ}$C of the same agar solution: $\mathcal{T}=1$~hour, 1 day, 2 days, 3 days, 4 days and 5~days. Experiments are performed on the hydrated gels. 
\label{fig.s2}}
\end{figure} 

\textbf{Supplemental Fig.~\ref{fig.s2}} shows the diffraction spectra of hydrated gels prepared after different incubation times $\mathcal{T}$ of the same agar solution, for $\mathcal{T}$ ranging from of a few hours to 5 days. Contrarily to the data reported in Fig.~5(d) in the main text of the manuscript which concern dried gels, the present experiments were  conducted on fully hydrated gels enclosed in sealed glass capillaries. The diffraction spectra show mainly a single maximum at $q_3=19.3~\mathrm{nm}^{-1}$ also visible on dry gels, and which amplitude decreases for increasing incubation times $\mathcal{T}$. The two other maxima visible on dry gels at $q_1=9.45~\mathrm{nm}^{-1}$ and $q_2=13.86~\mathrm{nm}^{-1}$ [see Fig.~5(d) in the main text] are barely visible here. Indeed water molecules in the gel, linked to the polymer network through hydrogen bonds are responsible for a modulation of the scattered peaks, thus degrading the angular spectrum resolution.

\begin{figure*}[t!]
\centering
\includegraphics[width=0.62\linewidth]{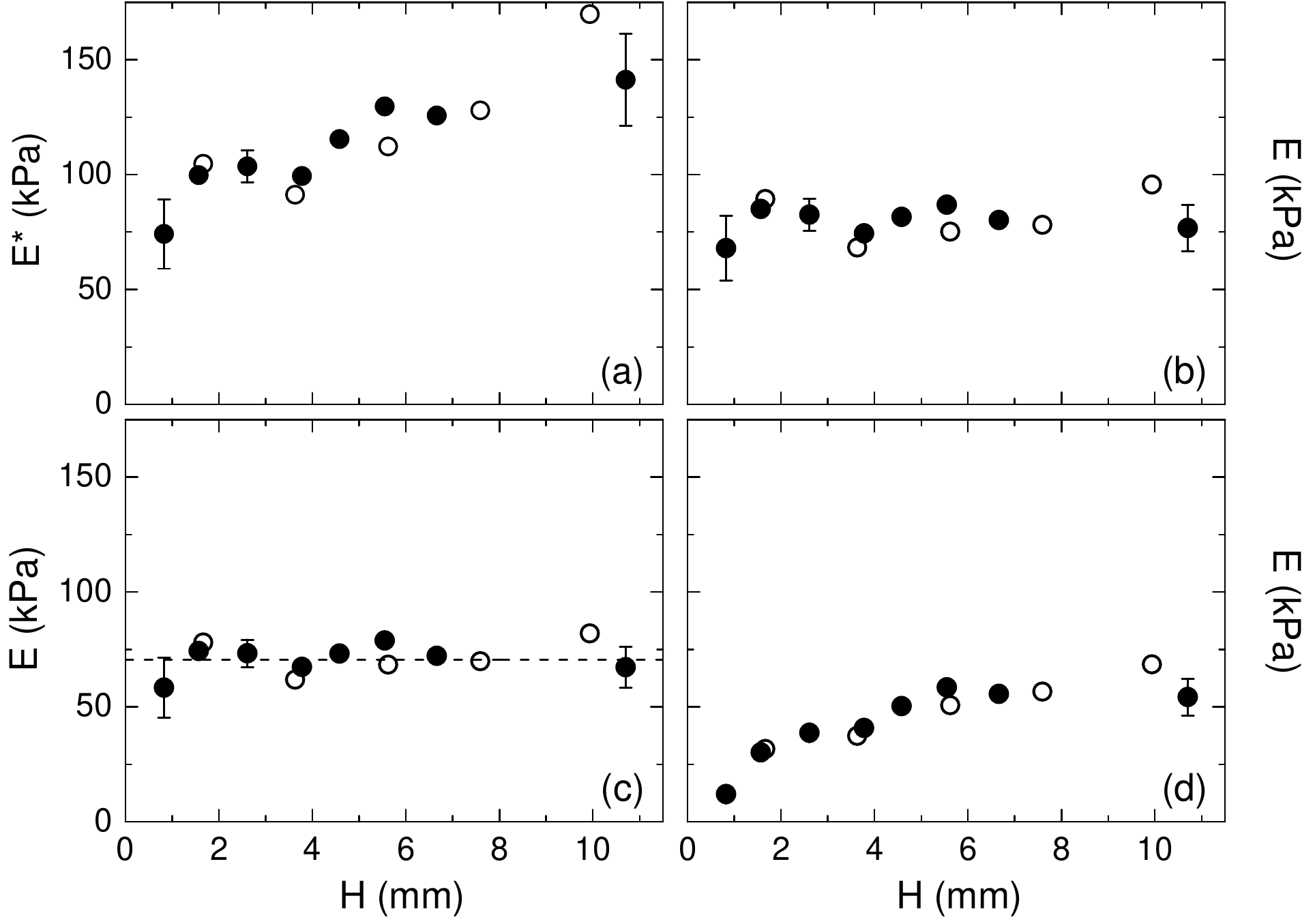}
\caption{(a) Evolution of the apparent compression modulus $E^* = \partial \sigma / \partial \epsilon$ as determined during indentation experiments vs the gel thickness $H$. The two types of symbols correspond to experiments performed at different indentation velocities: ($\circ$) $v=50$~$\mu$m/s and ($\bullet$) $v=100$~$\mu$m/s. Each point corresponds to a gel prepared from a fresh agar sol, i.e. from an agar solution incubated at $T=80^{\circ}$C for less than a day. (b)--(d) Compression modulus E derived from $E^*$ shown in (a) using the expression~(\ref{Modulus}) with a Poisson coefficient $\nu=0.0$ in (b), $\nu=0.3$ in (c) and $\nu=0.5$ in (d). The black horizontal dashed line in (c) corresponds to the best linear fit of the data, leading to $E=70\pm 7$kPa 
\label{fig.s3}}
\end{figure*} 

\textbf{Supplemental Fig.~\ref{fig.s3}(a)} shows the apparent elastic modulus $E^*$ as determined by the slope of the stress-strain relation $\sigma(\epsilon)$ at strains lower than a few percents for various gel thicknesses $H$. The former physical quantity is measured by macro-indentation experiments similar to the one reported in Fig.~8 in the main text, and performed at constant indentation velocity using a flat-ended cylinder (see subsection~2.2.1 in the main text for technical details). To convert the apparent compression modulus $E^*$ which increases with the gel thickness, into a true compression modulus $E$, we take into account the cylindrical shape of the indenter, the finite thickness $H$ of the gel disk which is of comparable size to the indenter diameter $2r=10$~mm and the stress singularity at the sharp edge of the indenter. 
\begin{figure}[!bh]
\centering
\includegraphics[width=\linewidth]{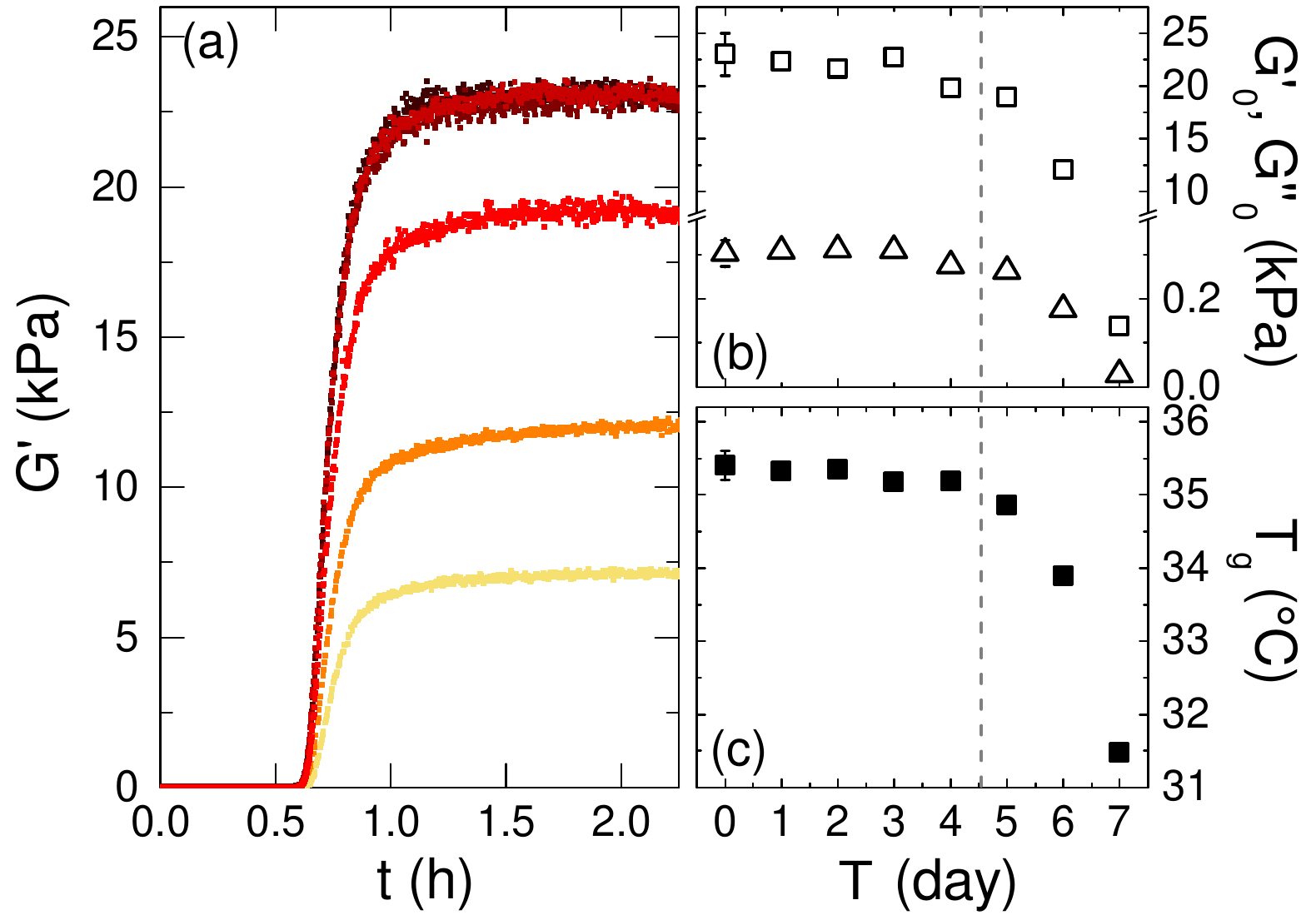}
\caption{ (Color online) Gelation experiments performed on a 1.5\% wt. agar sol, which agar is provided by BioM\'erieux instead of Sigma-Aldrich. (a) Elastic modulus $G'$ vs time $t$ during the cooling from $T=70^{\circ}$C down to 20$^{\circ}$C, at $\dot{\rm T}=1^{\circ}$C/min of various samples extracted from the same agar sol after various incubation times $\mathcal{T}=1$~hour, 2 days, 4, 6 and 7 days coded in color from black to yellow. (b) Steady-state values of the elastic ($G'$, $\square$) and viscous ($G''$, $\triangle$) moduli reached at the end of the gelation process vs $\mathcal{T}$. (c) Gelation temperature $T_g$ defined as the crossing temperature of $G'$ and $G''$, vs the incubation time $\mathcal{T}$. Error bars in (b) and (c) are only indicated on the first point and were determined by repeating the experiment with three different agar solutions. 
\label{fig.s5}}
\end{figure} 
The true compression modulus reads as follows \cite{Hayes:1972}:   
\begin{equation}
E=(1-\nu^2)\frac{r}{H}\frac{\pi}{2\kappa}E^*
\label{Modulus}
\end{equation}
where $\nu$ denotes the Poisson coefficient of the gel, and $\kappa$ is a numerical correction related to the finite size of the sample which depends on $r$, $H$ and $\nu$. The Poisson coefficient $\nu$ is unknown and necessary to determine the value of $\kappa$. We report in Fig~\ref{fig.s3}(b)-(d) the compression modulus computed for three different values of $\nu$, namely $\nu=0.0$, 0.3 and 0.5. In each case, the value of $\kappa$ is determined using the integral expression proposed in ref. \cite{Haider:1997} for $\nu=0$, and in ref. \cite{Hayes:1972} for $\nu>0$. Relying on the assumption that the true compression modulus $E$ should neither depend on the gel thickness $H$, nor on the indentation speed, we can rule out the value $\nu=0.5$ [Fig.~\ref{fig.s3}(d)]. Moreover the data obtained with $\nu=0.0$ also exhibit a non-negligible dependence with $H$ [Fig.~\ref{fig.s3}(b)]. Finally, the lowest dependence of $E$ with the gel height is obtained with the following interval $\nu$: $0.1\leq \nu \leq 0.3$, as illustrated in Fig.~\ref{fig.s3}(c) for $\nu=0.3$. The latter range of values is considered in the main part of the manuscript.


\providecommand*{\mcitethebibliography}{\thebibliography}
\csname @ifundefined\endcsname{endmcitethebibliography}
{\let\endmcitethebibliography\endthebibliography}{}

 \end{document}